\documentclass[a4paper]{article}
\usepackage{color,graphicx}
\usepackage{amsmath}
\usepackage{amsthm}
\usepackage{amssymb}
\usepackage{amsfonts}
\RequirePackage{ifpdf} 
\ifpdf
\usepackage[pdftex,bookmarks=true,
		   bookmarksnumbered=false,
		   bookmarksopen=false,
		   colorlinks=true,
		   linkcolor=webred] {hyperref}
\definecolor{webgreen}{rgb}{0, 0.5, 0} 
\definecolor{webblue}{rgb}{0, 0, 0.5} 
\definecolor{webred}{rgb}{0.5, 0, 0}   
\else
\newcommand{\href}[2]{ #1 }
\fi
\title{Abnormal Netsukuku Domain Name Anarchy}
\author{http://netsukuku.freaknet.org\\AlpT (@freaknet.org)}
\begin{document}
\maketitle

\begin{abstract}
	In this document, we present the Abnormal Netsukuku Domain Name
	Anarchy system. ANDNA is the distributed, non hierarchical and
	decentralised system of hostname management used in the Netsukuku
	network.
\end{abstract}

\section{Preface}
\label{sec:preface}

We're assuming that you already know the basics of the Netsukuku topology. If not, read the
topology document first: \cite{ntktopology}.

\section{Introduction}
ANDNA is the distributed, non hierarchical and decentralised system of
hostname management in Netsukuku. It substitutes the DNS. 

The ANDNA database is scattered inside all Netsukuku network. In the worst
case, every node will have to use few hundred kilobytes of memory.

\section{ANDNA}
  
ANDNA is based on the principle that every hostname $h$ is a string of maximum
512 bytes. By taking the hash $H(h)$ of the hostname, we obtain a number.
We consider this number an IP and we call \emph{hash node} the node which owns
the same IP. Every hash node will keep a small portion of the distributed
database of ANDNA.\\

\begin{verbatim}            
    Node n
ip: 123.123.123.123
hash( hostname: "netsukuku" ) == 11.22.33.44
                                   ||
                                   ||
                                 Node i
                             ip: 11.22.33.44
                   {    [ Andna database of the node i ]    }
                   { hash("netsukuku") ---> 123.123.123.123 }
\end{verbatim}

The registration of the hostname $h$, by the node $n$ proceeds as follow:
\begin{enumerate}
	\item $n$ computes $i=H(h)$
	\item $n$ sends a registration request to $i$
	\item $i$ associates in its database $H(h)$ to the IP of $n$, which is
		$IP(n)$.
\end{enumerate}
The resolution of the hostname $h$ by the node $m$:
\begin{enumerate}
	\item $m$ computes $i=H(h)$
	\item $m$ sends a resolution query to $i$
	\item $i$ sends back to $m$ the IP associated to $H(h)$, that in this
		example is $IP(n)$.
\end{enumerate}

\subsection{Hash gnode}
It is highly probable that the hash $H(h)$ doesn't corresponds to any active
node: if there are a total of $2^30$ active nodes and we are using IP of 32
bits, then by taking the 32 bit hash of $h$ the probability is equal to
$25\%$. Moreover, even the actives node can also exit from the network.

For this reasons, the registration of a hostname will be kept by all the nodes
of a single gnodes. This ensures the distribution and the redundancy of the
database. The gnode associated to $H(h)$ is the \emph{hash gnode}, and it is
simply the gnode of level 1 of the hash node $H(h)$. We will denote the former
with $gH(h)$ and the latter with $nH(h)$. In symbols we can write: $nH(h) \in
gH(h)$.

\subsection{Approximated hash gnode}
Since even the hash gnodes cannot exist, an approximation strategy is
utilised: the nearest active gnode to the hash gnode becomes the
\emph{rounded hash gnode}, which will substitute in full the original (non
existent) hash gnode. The rounded hash gnode is denoted with $rgH(h)$.

Generally, when we are referring to the gnode, which has accepted a
registration, there is no difference between the two kind of gnodes, they are
always called hash gnodes.

\subsection{Hook}

When a node hooks to Netsukuku, it becomes automatically part of a hash gnode,
thus it will also perform the \emph{andna hook}.
By hooking, it will get from its rnodes all the caches and databases, which
are already distributed inside its hash gnode.

\subsection{Cryptographic ANDNA}
  
Before making a request to ANDNA, a node generates a couple of RSA keys,
i.e. a public one (\emph{pub key}) and a private (\emph{priv key}). The size of the
pub key will be limited due to reasons of space.
The request of a hostname made to ANDNA will be signed with the private key
and in the same request the public key will be attached.
In this way, the node will be able to certify the true identity of its
future requests.

\subsection{Hostnames limit}

The maximum number of hostnames, which can be registered is 256.
This limit is necessary to prevent the massive registration of hostnames,
which may lead to a overload of the ANDNA system. One its side effect is to
slow down the action of spammers, which registers thousands of common words to
resell them.

\subsubsection{Implementation of the limit}
The \emph{counter node} is a node with an ip equal to the hash of the public
key of the node, which has registered the hostname $h$. We'll denote the
counter node associated to the node $n$ with $cr(n)$.
Thus, for every node $n$, which has registered a hostname $h$, exists its
corresponding counter node $cr(n)$.\\

The counter node $cr(n)$ will keep the number of hostnames registered by
$n$.\\

When a hash gnode receives the registration request sent by $n$, it contacts
$cr(n)$, which replies by telling how many hostnames have been already registered
by the $n$. If the $n$ has not exceeded its limit, then $cr(n)$  will increment its counter
and will allow the hash gnode to register the hostname.

$cr(n)$ is activated by the check request the hash gnode sends. $n$ has to
keep the $cr(n)$ active following the same rules of the hibernation (see the
chapter below). Practically, if $cr(n)$ doesn't receives any more check
requests, it will deactivate itself, and all the registered hostnames will
become invalid.

The same distribution technique used for the hash gnode is used for the counter
node: there is a whole gnode of counter nodes, which is called, indeed,
counter gnode.

\subsection{Registration step by step} 
\begin{enumerate}
	\item The node $n$ wants to register its hostname $h$,
	\item it finds the rounded hash gnode $rgH(h)$ and contacts, randomly,
		one of its node. Let this node be $y$.
	\item $n$ sends the registration request to $y$. The request includes
		the public key of the node $n$. The pkt is also signed with
		the private key of $n$.
	\item The node $y$ verifies to be effectively the nearest gnode to the
		hash gnode, on the contrary it rejects the request.
		The signature validity is also checked.
	\item The node $y$ contacts the counter gnode $cr(n)$ and sends to it
		the IP, the hash of th hostname to be registered and a copy of
		the registration request itself.
	\item $cr(n)$ checks the data and gives its ok.
	\item The node $y$, after the affirmative reply, accepts the
		registration request and adds the entry in its database,
		storing the date of registration,
	\item $y$ forwards to its gnode the registration request.
	\item The nodes, which receive the forwarded request, will check its
		validity and store the entry in their db.
\end{enumerate}
  
\subsection{Hostname hibernation}

The hash gnode keeps the hostname in an hibernated state for about 30 days
since the moment of its registration or last update.
The expiration time is very long to stabilise the domains. In this way, even
someone attacks a node to steal its domain, it will have to wait 30 days
to fully have it.

When the expiration time runs out, all the expired hostnames are deleted and
substituted with the others in queue.\\
A node has to send an update request for each of its hostnames, each time it
changes ip and before the hibernation time expires, in this way it's
hostname won't be deleted.

The packet of the update request has an id, which is equal to the number of
updates already sent. The pkt is also signed with the private key of the
node to warrant the true identity of the request.
The pkt is sent to any node of the hash gnode, which will send a copy of the
request to the counter gnode, in order to verify if it is still active and
that the entry related to the hostname being updated exists. On the
contrary, the update request is rejected.

\subsection{Hash gnodes mutation}
 
If a generical rounded gnode is overpassed by a new gnode, which is nearer
to the hash gnode, it will exchange its role with that of the second one, i.e.
the new gnode will become the rounded hash gnode.\\

This transition takes place passively. When the register node will update
its hostname, it will directly contact the new rounded gnode. Since the
hostname stored inside the old rounded gnode won't be updated, it will be
dropped.\\

While the hostname has not been updated yet, all the nodes trying
to resolve it, will find the new rounded gnode as the gnode nearest to the
hash gnode and so they'll send their resolution queries to the new gnode.\\
When the  new rounded gnode receives the first query, it will notice that it
is indeed a rounded gnode and to reply to the query, it it will retrieve from
the old hash gnode the andna cache related to the hostname to resolve.\\
In this way, the registration of that hostname is automatically transfered
from the old gnode into the new one.\\

If an attacker is able to send a registration request to the new hash gnode,
before that the rightful owner of the hostname updates it or before that the
passive transfer occurs, it will be able to steal it.\\
To prevent this, the new hash gnode will always double check a registration
request by contacting the old hash gnode.

\subsection{Yet another queue}
  
Every node is free to choose any hostname, even if the hostname has been
already chosen by another node.\\
The andna cache will keep a queue of MAX\_ANDNA\_QUEUE (5) entries.
When the hostname on top of the queue expires, it will be automatically
substituted by the next hostname.

\subsection{Distributed cache for hostname resolution}

In order to optimise the resolution of a hostname, a simple strategy is
used: a node, each time it resolves a hostname, stores the result in a
cache. For each next resolution of the same hostname, the node has already
the result in its cache.\\
Since the resolution packet contains the latest time when the hostname has
been registered or updated, an entry in the cache will
expires exactly at the same time of the ANDNA hostname expiration.\\

The \emph{resolved hnames} cache is readable by any node.\\
A node $x$, exploiting this feature, can ask to any node $y$, randomly chosen
inside its same gnode, to resolve for itself the given hostname.
The node $y$, will search in its resolved cache the hostname and on negative
result the node will resolve it in the standard way, sending the result to
the node $x$.
These technique avoids the overload of the hash gnodes keeping very famous
hostnames.

\subsection{noituloser emantsoh esreveR}

If a node wants to know all the hostnames associated to an ip, it will
directly contact the node which owns that ip.

\subsection{DNS wrapper}

The DNS wrapper listen to the resolution queries sent by localhost and wraps
them: it sends to the ANDNA daemon the hostnames to resolve and receives 
the their resolutions.\\

Thanks to the wrapper it is possible to use the ANDNA without modifying
any preexistent programs.\\
  
See the ANDNS RFC \cite{ANDNSRFC} and the ANDNA manual \cite{manandna}.

\section{Scattered Name Service Disgregation}
The updated SNSD RFC can be found here: \cite{snsd}\\

The Scattered Name Service Disgregation is the ANDNA equivalent of the
SRV Record\cite{SRV} of the Internet Domain Name System.\\

SNSD isn't the same of the ``SRV Record'', in fact, it has its own unique
features.\\
  
With the SNSD it is possible to associate IPs and hostnames to another
hostname.

Each assigned record has a service number, in this way the IPs and hostnames
which have the same service number are grouped in an array.
In the resolution request the client will specify the service number too,
therefore it will get the record of the specified service number which is 
associated to the hostname. Example:
\begin{enumerate}
	\item The node X has registered the hostname "angelica". The default
		IP of ``angelica'' is ``1.2.3.4''.
	\item  X associates the ``depausceve'' hostname to the `http' service
		number (80) of ``angelica''.
	\item  X associates the ``11.22.33.44'' IP to the `ftp' service number (21) of
  ``angelica''.
\end{enumerate}
  
When the node Y resolves normally ``angelica'', it gets 1.2.3.4, but when
its web browser tries to resolve it, it asks for the record associated to
the `http' service, therefore the resolution will return "depausceve".
The browser will resolve "depausceve" and will finally contact the server.
When the ftp client of Y will try to resolve "angelica", it will get the
"11.22.33.44" IP.\\
  
The node associated to a SNSD record is called "SNSD node". In this example
"depausceve" and 11.22.33.44 are SNSD nodes.\\

The node which registers the records and keeps the registration of the main
hostname is always called "register node", but it can also be named "Zero SNSD
node", in fact, it corresponds to the most general SNSD record: the service
number 0.\\
  
Note that with the SNSD, the NTK\_RFC 0004 will be completely deprecated.

\subsection{Service, priority and weight number}

\subsubsection{Service number}
  
The service number specifies the scope of a SNSD record. The IP associated to 
the service number `x' will be returned only to a resolution request which has
the same service number.\\
  
A service number is the port number of a specific service. The port of the
service can be retrieved from /etc/services.\\
  
The service number 0 corresponds to a normal ANDNA record. The relative IP
will be returned to a general resolution request.\\
  
\subsubsection{Priority}
  
The SNSD record has also a priority number. This number specifies the priority
of the record inside its service array.\\
The client will contact first the SNSD nodes which have the higher priority,
and only if they are unreachable, it will try to contact the other nodes
which have a lower priority.
  
\subsubsection{Weight}
  
The weight number, associated to each SNSD record, is used when there are
more than one records which have the same priority number.
In this case, this is how the client chooses which record using to contact
the servers:\\
  
The client asks ANDNA the resolution request and it gets, for example, 8
different records.\\
The first record which will be used by the client is chosen in a pseudo-random
manner: each record has a probability to be picked, which is proportional to its
weight number, therefore the records with the heavier weight are more likely to
be picked.\\
Note that if the records have the same priority, then the choice is completely
random.\\

It is also possible to use a weight equal to zero to disable a record.\\

The weight number has to be less than 128.\\

\subsection{SNSD Registration}

The registration method of a SNSD record is similar to that described in the
NTK\_RFC 0004.\\

It is possible to associate up to 16 records to a single hostname.
The maximum number of total records which can be registered is 256.\\

The registration of the SNSD records is performed by the same register node.
The hash node which receives the registration won't contact the counter node,
because the hostname is already registered and it doesn't need to verify
anything about it. It has only to check the validity of the signature.\\

The register node can also choose to use an optional SNSD feature to be sure
that a SNSD hostname is always associated to its trusted machine. In this
case, the register node needs the ANDNA pubkey of the SNSD node to send a
periodical challenge to the node.
If the node fails to reply, the register node will send to ANDNA a delete
request for the relative SNSD record.\\

The registration of SNSD records of hostnames which are only queued in the
andna queue is discarded.

Practically, the steps necessary to register a SNSD record are:
\begin{enumerate}
	\item Modify the \verb|/etc/netsukuku/snsd_nodes| file.
\begin{verbatim}
register_node# cd /etc/netsukuku/ 
register_node# cat snsd_nodes
#
# SNSD nodes file
#
# The format is:
# hostname:snsd_hostname:service:priority:weight[:pub_key_file]
# or
# hostname:snsd_ip:service:priority:weight[:pub_key_file]
#
# The `pub_key_file' parameter is optional. If you specify it, NetsukukuD will
# check periodically `snsd_hostname' and it will verify if it is always the
# same machine. If it isn't, the relative snsd will be deleted.
#

depausceve:pippo:http:1
depausceve:1.2.3.4:21:0

angelica:frenzu:ssh:1:/etc/netsukuku/snsd/frenzu.pubk

register_node#
register_node# scp frenzu:/usr/share/andna_lcl_keyring snsd/frenzu.pubk
\end{verbatim}
\item Send a SIGHUP to the NetsukukuD of the register node:
\begin{verbatim}
register_node# killall -HUP ntkd
# or, alternatively
register_node# rc.ntk reload
\end{verbatim}
\end{enumerate}

\subsubsection{Zero SNSD IP}

The main IP associated to a normal hostname has these default values:
\begin{verbatim}
IP	 = register_node IP	# This value can't be changed
service  = 0
priority = 16
weight   = 1
\end{verbatim}

It is possible to associate other SNSD records in the service 0, but it isn't
allowed to change the main IP. The main IP can only be the IP of the
register\_node.\\
Although it isn't possible to set a different association for the main IP, it
can be disabled by setting its weight number to 0.\\

The string used to change the priority and weight value of the main IP is:
\begin{verbatim}
hostname:hostname:0:priority:weight

# For example:
register_node# echo depausceve:depausceve:0:23:12 >> /etc/netsukuku/snsd_nodes
\end{verbatim}

\subsubsection{SNSD chain}

Since it is possible to assign different aliases and backup IPs to the zero
record, there is the possibility to create a SNSD chain.
For example:

\begin{verbatim}
depausceve registers: depausceve:80 --> pippo
pippo registers:      pippo:0  --> frenzu
frenzu registers:     frenzu:0 --> angelica
\end{verbatim}

However the SNSD chains are ignored, only the first resolution is considered
valid. Since in the zero service there's always the main IP, the resolution is
always performed.\\
In this case ("depausceve:80 --> pippo:0") the resolution will return the main
IP of "pippo:0".\\

The reply to a resolution request of service zero, returns always IPs and not
hostnames.


\newpage

\begin{center}
\verb|^_^|
\end{center}

\end{document}